\documentclass[10pt,aps,prd,twocolumn,a4paper,showpacs,nofootinbib]{revtex4-1}
\usepackage{amsmath}
\usepackage{amsfonts}
\usepackage[colorlinks=true,linkcolor=blue,citecolor=blue,urlcolor=blue]{hyperref}
\usepackage{amssymb}
\usepackage{graphicx}
\usepackage{enumerate}

\providecommand{\abs}[1]{\left|#1\right|}
\providecommand{\ep}[1]{{e}^{#1}}
\newcommand{\bk}{\mathbf{k}}

\begin{document}

\title{Quantum entanglement due to modulated Dynamical Casimir Effect} 
\author{Xavier Busch}\email[]{xavier.busch@th.u-psud.fr}
\author{Renaud Parentani}\email[]{renaud.parentani@th.u-psud.fr}
\author{Scott Robertson}\email[]{scott.robertson@th.u-psud.fr}
\affiliation{Laboratoire de Physique Th\'eorique, CNRS UMR 8627, B{\^{a}}timent\ 210,
         \\Universit\'e Paris-Sud 11, 91405 Orsay CEDEX, France}


\begin{abstract}

We study the creation and entanglement of quasiparticle pairs due to a periodic variation of the mode frequencies of a homogeneous quantum system. Depending on the values of the parameters describing the periodic modulation, the number of created pairs either oscillates or, in a narrow resonant frequency interval, grows exponentially in time. For a system initially in a thermal state, we determine in which cases the final state is quantum mechanically entangled, i.e., where the bipartite state is nonseparable. We include some weak dissipation, expected to be found in any experimental setup, and study the corresponding reduction of the quantum entanglement. Our findings are used to interpret the results of two recent experiments.

\end{abstract}

\maketitle

\section{Introduction}
It is well known that a sudden change occurring in a system, be it an expanding universe~\cite{Parker:1975jm,Starobinsky:1979ty,Mukhanov:1981xt,Kofman:1997yn}, a Bose-Einstein condensate~\cite{Fedichev:2003bv,Carusotto:2009re}, or a superconducting circuit~\cite{WilsonDCE,Lahteenmaki12022013,Felicetti:arXiv1402.4451}, induces a parametric amplification of linear density perturbations. In quantum settings, this amplification implies the production of pairs of (quasi-)particles, and if the system is homogeneous the two particles have opposite wave vectors $\bk$ so that the total momentum is conserved. It is also known that these pairs are maximally entangled when the system is initially in its vacuum state~\cite{Campo:2005sv}. If instead the initial state is populated in an incoherent manner -- as is the case in a thermal bath -- the stimulated effects tend to reduce the quantum entanglement~\cite{Bruschi:2013tza,Busch:2013sma,finazzi2013}. A similar reduction in coherence occurs when the system is coupled to an external environment~\cite{Campo:2005sy,Campo:2008ju,Adamek:2013vw,Busch:2013sma,Busch:2013gna}.

The loss of quantum entanglement can be precisely characterized by using the mean occupation number $n_k \doteq {\rm Tr} [ \hat \rho \hat b_\bk^\dagger \hat b_\bk ] $ and the complex correlation term $c_k \doteq {\rm Tr} [ \hat \rho \hat b_{-\bk} \hat b_\bk ] $, where $\hat \rho$ is the density operator of the system and $ \hat b_\bk^\dagger$ and $\hat b_\bk $ are, respectively, the creation and destruction operators of a quasiparticle of momentum $\bk$. In fact, whenever 
\begin{equation}
\Delta_k \doteq n_k- \abs{c_k} , 
\label{Ineq}
\end{equation}
is negative, the state of the bipartite $\bk, - \bk$ system is said to be \textit{nonseparable}~\cite{Werner:1989,Simon:2000zz}: The quasiparticles are so strongly correlated that the statistical properties of the system cannot be described by a classical stochastic ensemble. See Refs.~\cite{Campo:2005sy,Adamek:2013vw,Busch:2013sma} for a presentation and use of these concepts. As explained in these references, the criterion $\Delta_k < 0$ is sufficient to get nonseparability in the general case. It is also a necessary condition for homogeneous, isotropic, and Gaussian states.

While entanglement and nonseparability are well established in theory, it is only recently that they have become experimentally accessible in the present context of pair creation due to a temporal change. The main difficulties arise from the weakness of the pair creation rate, and in making sufficiently precise measurements of $\abs{c_k}$ and $n_k$ to be able to assert nonseparability. There have been two recent experiments able to test the condition $\Delta_k < 0$: first using an atomic BEC~\cite{PhysRevLett.109.220401} and second, in a Josephson metamaterial~\cite{Lahteenmaki12022013}. Interestingly, in the first case, the condition was not met, while it was met in the second. In both cases, the experimental teams used a periodic modulation of some parameter over an extended period of time. (See also Ref.~\cite{WilsonDCE} for an earlier experiment, and Ref.~\cite{Carusotto:2009re} for an earlier theoretical study of such a modulation.) The advantage of this method over a single sudden change is that, even though the creation rate may be small, it could conceivably engender a very large pair production if the modulation lasts long enough. Indeed, when the frequency $\omega_{\bk}$ of a pair is close to half of the modulation frequency $\omega_p/2$, the number of such created pairs increases exponentially with the duration of the modulation (for bosonic excitations).

In this paper we study theoretically the reaction of a system to a long-lasting modulation with the aim of understanding under which conditions the final state will be nonseparable. As in the case of a sudden change~\cite{Bruschi:2013tza,Busch:2013sma}, the sign of the parameter $\Delta_k$ of Eq.~\eqref{Ineq} is fixed by only two quantities: $\abs{\beta^{\rm as}_k}$, the magnitude of the asymptotic Bogoliubov coefficient which governs the spontaneous effects and reduces the value of $\Delta_k$, and the initial mean occupation number of incoherent quanta $n_{k, \rm in}$ which governs the induced effects and increases $\Delta_k$. The behavior of $\abs{\beta^{\rm{as}}_k}$ is non trivial, its value being fixed in a convoluted way by several parameters describing the modulation.  As far as we know, the aspects of the system associated to the behavior of $\abs{\beta^{\rm{as}}_k}$ are as yet unstudied.

Our paper is organized as follows. In Sec.~\ref{sec:tempchangeeom} we present the basic equations governing the time evolution of the modes of a homogeneous system in response to an arbitrary modulation in time; this response is characterized by the time dependent Bogoliubov coefficients, from which the mean occupation number and the separability parameter of Eq.~\eqref{Ineq} are derived. In Sec.~\ref{sec:numanalisys} we introduce a particular modulation, i.e., a sinusoidal oscillation of the mode frequencies, and the parameters that describe it, before a numerical investigation of the dependence of the final state of the system (its particle content and its degree of entanglement) on these parameters. (The properties of the Bogoliubov coefficients observed in Sec.~\ref{sec:numanalisys} are derived analytically in Appendix~\ref{app:analytic}.) In Sec.~\ref{sec:dissip} (with further details in Appendix~\ref{app:dissip}) we show how to incorporate weakly dissipative effects into the picture. The paper concludes with Sec.~\ref{sec:conclusion}.

\section{Time-dependence in homogeneous media}
\label{sec:tempchangeeom}

In this section we consider the effects of time dependence on a quantum system. While the nature of the time dependence is left unspecified, we shall restrict our attention to homogeneous media, allowing the entire analysis to be done at fixed wave vector $\bk$. As a result, the dimensionality of the system drops out, and need not be specified. To fix the notation and the concepts, we shall work in an atomic Bose condensate~\cite{Dalfovo:1999zz,PhysRevLett.109.220401}. However, the following analysis is easily adapted to other media, such as polariton systems~\cite{Carusotto:2012vz} and Josephson metamaterials~\cite{WilsonDCE,Lahteenmaki12022013}. It is also applicable to pair creation in cosmological models, such as primordial inflation~\cite{Starobinsky:1979ty,Mukhanov:1981xt}; in particular, the time variation we shall study in Sec.~\ref{sec:numanalisys} is very similar to that occurring during the preheating phase at the end of inflation~\cite{Kofman:1997yn}. 

\subsection{Equations of motion}

In a condensed dilute gas, linear density perturbations obey the Bogoliubov-de~Gennes equation~\cite{Dalfovo:1999zz}. At fixed $\bk$, in units where $\hbar = 1$, one obtains (see, e.g.,~\cite{Busch:2013sma}) 
\begin{equation}
\label{eq:eomphi}
i\partial_t \hat \phi_\bk = \Omega_k \hat \phi_\bk + m c^2 \hat \phi_{-\bk}^\dagger ,
\end{equation}
where $m$ is the atom mass, $\Omega_k \doteq \frac{k^2}{2m} +m c^2 $, and $c$ is the low-frequency speed of sound\footnote{
In a homogeneous polariton system with coherent pumping, one obtains a similar equation for quasiparticle excitations of the lower branch with $\Omega_k = \frac{k^2}{2m} - \omega_p + E_0 + 2 m c^2$, where $m$ is the effective mass of the photons, $\omega_p$ is the pump frequency, and $E_0$ is the rest energy (see Ref.~\cite{Busch:2013gna}).}. As in Refs.~\cite{PhysRevA.67.033602,Finazzi:2010nc}, we shall describe Eq.~\eqref{eq:eomphi}, as well as its corresponding Hermitian conjugate equation with $\bk \to -\bk$, as a matrix equation:
\begin{equation}
\begin{split}
i\partial_t \left [ \begin{array}{ll}
 \hat \phi_\bk\\
\hat \phi_{-\bk}^\dagger
\end{array}\right ] &= \left [ \begin{array}{ll}
 \Omega_k & mc^2 \\
-mc^2 & -\Omega_k
\end{array}\right ] \times
\left [ \begin{array}{ll}
 \hat \phi_\bk\\
\hat \phi_{-\bk}^\dagger
\end{array}\right ].
\end{split}
\end{equation}
To clearly identify the effects that are due to a temporal change of $\Omega_k$ or $c$, we perform the standard Bogoliubov transformation 
\begin{equation}
\label{eq:defbogofield}
\begin{split}
\left [ \begin{array}{ll}
 \hat \varphi_\bk\\
\hat \varphi_{-\bk}^\dagger
\end{array}\right ] &= \left [ \begin{array}{ll}
u_k & v_k\\
v_k & u_k
\end{array}\right ]\times \left [ \begin{array}{ll}
 \hat \phi_\bk\\
\hat \phi_{-\bk}^\dagger
\end{array}\right ] ,
\end{split}
\end{equation}
where $u_k$ and $v_k$ are given by
\begin{equation}
\begin{split}
u_k &\doteq \frac{\sqrt{\Omega_k+mc^2}+ \sqrt{\Omega_k-mc^2}}{2\sqrt{\omega_k}}, \\
v_k &\doteq \frac{\sqrt{\Omega_k+mc^2}- \sqrt{\Omega_k-mc^2}}{2\sqrt{\omega_k}}
\end{split}
\label{eq:ukandvk}
\end{equation} 
and the frequency $\omega_k$ is given by
\begin{equation}
\begin{split}
\omega_k^2 \doteq \Omega_k^2 - m^2 c^4 .
\end{split}
\label{eq:omegak}
\end{equation}
When $\Omega_k$ and/or $c$ vary in time, so too do $u_k$, $v_k$, and $\omega_k$ via Eqs.~\eqref{eq:ukandvk} and~\eqref{eq:omegak}. Using the fact that $u_k^2-v_k^2=1$, straightforward algebraic manipulation leads to the following equation of motion\footnote{
Equation~\eqref{eq:eomvarphi} is very similar to the equation governing the photon field in a cavity of modulated Josephson metamaterial in Ref.~\cite{Lahteenmaki12022013}, although due to the stationarity of that inhomogeneous system (in a rotating frame), the correlations are between opposite frequencies rather than opposite wave vectors.}
for the Bogoliubov operators\footnote{Note that, as in Ref.~\cite{Busch:2013sma}, we could also have considered $\hat{\chi}_{\bk} \propto \hat{\phi}_{\bk}+\hat{\phi}_{-\bk}^{\dagger}$ which obeys $\left[\partial_{t}^{2}+\omega_{k}^2(t)\right]\hat{\chi}_{\bk}=0$. For a sinusoidal modulation of $\omega_{k}^{2}(t)$, this is (up to a coordinate transformation) the Mathieu equation~\cite{Abramowitz}, which also plays a role in preheating cosmological scenarios~\cite{Kofman:1997yn}. } $\varphi_\bk$ and $\hat \varphi_{-\bk}^\dagger$:
\begin{equation}
\label{eq:eomvarphi}
\begin{split}
i\partial_t \left [ \begin{array}{ll}
 \hat \varphi_\bk\\
 \hat \varphi_{-\bk}^\dagger
\end{array}\right ] &= \left [ \begin{array}{ll}
 \omega_k & i \frac{\dot u_k}{v_k} \\
i \frac{\dot u_k}{v_k} 
& - \omega_k
\end{array}\right ] \times
\left [ \begin{array}{ll}
 \hat \varphi_\bk\\
 \hat \varphi_{-\bk}^\dagger
\end{array}\right ] ,
\end{split}
\end{equation}
where $\dot u_k = \partial_t u_k$.

In stationary systems, one recovers the standard diagonal matrix governed by $\omega_k$. In that case, the fields are trivially related to the (canonical) phonon creation and annihilation operators $\hat b_{-\bk}^\dagger$ and $\hat b_\bk$, obeying $[\hat b_\bk,(\hat b_{\bk'})^\dagger] = \delta(\bk-\bk')$, by 
\begin{equation}
\label{eq:statiovarphi}
\begin{split}
\left [ \begin{array}{ll}
\hat \varphi_\bk\\
 \hat \varphi_{-\bk}^\dagger
\end{array}\right ]
&= \left [ \begin{array}{ll}
\ep{- i \omega_k t}\\
0
\end{array}\right ] \hat b_\bk + \left [ \begin{array}{ll}
0\\
\ep{ i \omega_k t}
\end{array}\right ](\hat b_{-\bk})^\dagger .
\end{split}
\end{equation}
When the system is stationary for asymptotic early times, the initial operators $\hat b_\bk^{\rm in}$ and $(b_{-\bk}^{\rm in})^\dagger$ are well defined and related at early times to the field operators by the above equation. The same is true when the system becomes stationary for asymptotic late times, where the late behavior of the fields $\varphi_\bk$ and $\hat \varphi_{-\bk}^\dagger$ defines the final operators $\hat b_\bk^{\rm out}$ and $(b_{-\bk}^{\rm out})^\dagger$. Then, because the field equation is linear, the two sets of asymptotic operators are related by an overall Bogoliubov transformation
\begin{equation}
\label{eq:bogoliubov}
\begin{split}
\hat b_\bk^{\rm out} = \alpha_k^{\rm as} \hat b_\bk^{\rm in} + ( \beta_k^{\rm as})^* (\hat b_{-\bk}^{\rm in})^\dagger ,
\end{split}
\end{equation}
where the requirement that both the initial and final operators satisfy the bosonic commutation relations imposes the condition $\abs{\alpha_k^{\rm as}}^{2} - \abs{\beta_k^{\rm as}}^{2} = 1$.

The asymptotic $in$ operators define a two-component mode $W_k^{\rm in}(t)$ via the commutator
\begin{equation}
 W_k^{\rm in}(t) \doteq \left [ 
 \begin{array}{ll}
\left [\hat \varphi_\bk(t), (b^{\rm in}_\bk)^\dagger\right ]  \\  
 \null [\hat \varphi^{\dagger}_{-\bk}(t) , (b^{\rm in}_\bk)^\dagger  ]
\end{array} \right ] .
\label{phimode}
\end{equation}
To simplify the notation, we shall no longer write the subscript $\bk$ since all equations shall be defined for a fixed value of $k = \abs{\bk}$. Equation~\eqref{phimode} implies that the mode doublet $W^{\rm in}(t)$ is the solution of Eq.~\eqref{eq:eomvarphi} with initial conditions $W^{\rm in}\underset{t \to -\infty}\sim \left [ \begin{array}{ll}
\ep{- i \omega t}\\
0
\end{array}\right ]$. 
For large times it behaves as 
\begin{equation}
\begin{split}
W^{\rm in} \underset{t \to +\infty}\sim \left [ \begin{array}{ll}
\alpha^{\rm as} \ep{- i \omega t} \\
\beta^{\rm as} \ep{ i \omega t} 
\end{array}\right ].
\end{split}
\end{equation}
Following the standard method~\cite{Massar:1997en} to evaluate the Bogoliubov coefficients $\alpha^{\rm as}$ and $\beta^{\rm as}$, we introduce the functions $\alpha(t)$ and $\beta(t)$ through the expression
\begin{equation}
\begin{split}
W^{\rm in}(t) = \left [ \begin{array}{ll}
\alpha(t) \, \ep{- i \int^t \omega dt'} \\
\beta(t) \, \ep{ i \int^t \omega dt'} 
\end{array}\right ] .
\end{split}
\end{equation}
By definition, their initial values are $1$ and $0$, and their late-time values coincide (up to a phase) with $\alpha^{\rm as}$ and $\beta^{\rm as}$. They obey the first-order coupled equations 
\begin{equation}
\label{eq:eomalphabeta}
\begin{split}
\partial_t \alpha &= \frac{\dot u}{v} \ep{ 2i \int \omega dt} \beta , \\
\partial_t \beta &= \frac{\dot u}{v} \ep{- 2i \int \omega dt} \alpha .
\end{split}
\end{equation}
One then verifies that the zeroth order solution, i.e., constant values for $\alpha$ and $\beta$, corresponds to the WKB approximation. One also verifies that corrections are associated with nonadiabatic transitions, and are here interpreted as creation of phonon pairs with opposite wave vectors.

It is interesting to notice that
\begin{equation}
\begin{split}
\frac{\dot u }{v}= \frac{\dot \omega }{ 2\omega} - \frac{\partial_t(\Omega - mc^2)}{2(\Omega - mc^2)}.
\end{split}
\end{equation}
In the following, we shall assume that $\Omega - mc^2$ is constant, or equivalently that the mass $m$ is constant.\footnote{Note that this is not necessarily true in some systems, such as polaritons or atoms on a lattice, as pointed out to us by C.~Westbrook~\cite{RevModPhys.78.179}.} In this case, the Bogoliubov coefficients are governed solely by the time evolution of $\omega$.

\subsection{Occupation number and nonseparability}

When the initial quasiparticle state is incoherent, homogeneous, isotropic, and Gaussian, it is fully governed by the mean occupation number $n_{\rm in}$, since the initial coherence $c_{\rm in}$ vanishes. When the time dependent perturbation preserves the homogeneity, the final state is still Gaussian, homogeneous, and fully governed by
\begin{subequations}
\label{eq:outnc} 
\begin{align}
n_{\rm out} &=\abs{\beta^{\rm as}}^2 + n_{\rm in}\left[\abs{\alpha^{\rm as}}^2 + \abs{\beta^{\rm as}}^2 \right], \\
c_{\rm out} &= \alpha^{\rm as} \left(\beta^{\rm as}\right)^{*} + n_{\rm in} \left [2 \alpha^{\rm as} \left(\beta^{\rm as}\right)^{*} \right ]. 
\end{align}
\end{subequations}
The first terms of these equations give the contributions of spontaneous effects, whereas the second, proportional to $n_{\rm in}$, arise from induced effects.

As explained in the Introduction, the final state is nonseparable whenever $n_{\rm out} - \abs{c_{\rm out}} < 0$. Equations~\eqref{eq:outnc} imply that the final value of $\Delta$ of Eq.~\eqref{Ineq} is given by~\cite{Busch:2013sma}
\begin{equation}
\label{eq:deltaout}
\begin{split}
\Delta_{\rm out} &= -(\abs{\alpha^{\rm as}}- \abs{\beta^{\rm as}}) \abs{\beta^{\rm as}} + n_{\rm in}\left (\abs{\alpha^{\rm as}} - \abs{\beta^{\rm as}}\right )^2 .
\end{split}
\end{equation}
From this equation it is clear that increasing $n_{\rm in}$ necessarily increases the value of $\Delta_{\rm out}$, thereby establishing that induced effects cannot give rise to nonseparable states. However, Eq.~\eqref{eq:deltaout} does not indicate what results when increasing the value of $\abs{\beta^{\rm as}}$. To reveal its effect, we use $\abs{\alpha^{\rm as}}^{2} - \abs{\beta^{\rm as}}^{2} = 1$, and rewrite $\Delta_{\rm out}$ as
\begin{equation}
\label{eq:deltaoutbeta}
\begin{split}
\Delta_{\rm out} &= \frac{-\abs{\beta^{\rm as}}\left(\abs{\beta^{\rm as}} + \abs{\alpha^{\rm as}}\right) + n_{\rm in}}{\left(\abs{\beta^{\rm as}} + \abs{\alpha^{\rm as}}\right)^{2}} .
\end{split}
\end{equation}
From this it can be verified that $\Delta_{\rm out}$ strictly decreases with increasing $\abs{\beta^{\rm as}}$, and that $\Delta_{\rm out}<0$ when $\left(\abs{\beta^{\rm as}}+\abs{\alpha^{\rm as}}\right)^{2} > 2n^{\rm in}+1$. This will play a crucial role in what follows because in a certain regime, we shall see that $\abs{\beta^{\rm as}} $ exponentially increases in time, thereby giving rise to nonseparable states when the modulation of $\omega$ lasts long enough. 

\section{Numerical analysis} 
\label{sec:numanalisys}

\begin{figure}[b]
\includegraphics[width=1\linewidth]{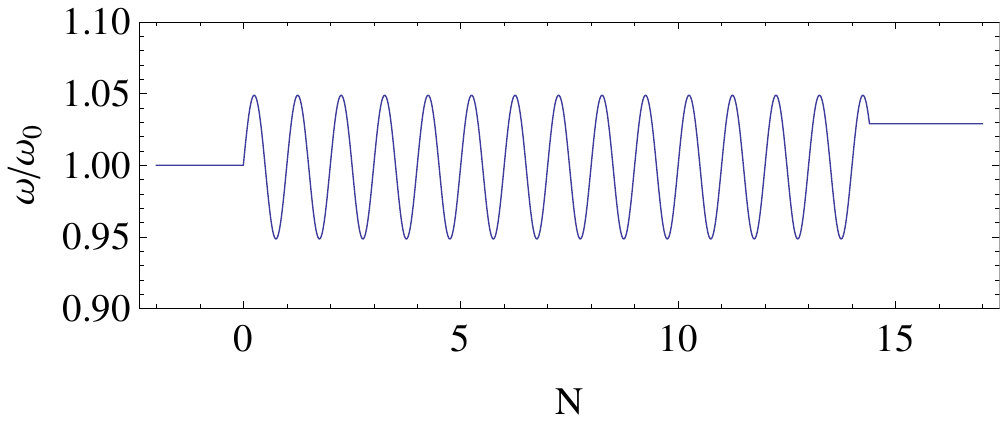}
\caption{Here is plotted an example of the frequency modulation of Eq.~\eqref{modul}, square-rooted so as to give $\omega/\omega_0$. The horizontal axis represents $N =\omega_p t/2\pi$. The amplitude $A=0.1$ (which applies to the squared frequency, being replaced by approximately $A/2$ when the square root is taken), and the total number of oscillations $N=14.4$.}
\label{fig:omegaoft}
\end{figure}

Here we apply the concepts described in Sec.~\ref{sec:tempchangeeom} to a specific type of temporal modulation, numerically solving Eqs.~\eqref{eq:eomalphabeta} to find the behavior of the Bogoliubov coefficient $\abs{\beta^{\rm as}}$ and of the parameter $\Delta_{\rm out}$. 

\subsection{Frequency modulation}

\begin{figure}
\begin{minipage}{1\linewidth}
\includegraphics[width=1\linewidth]{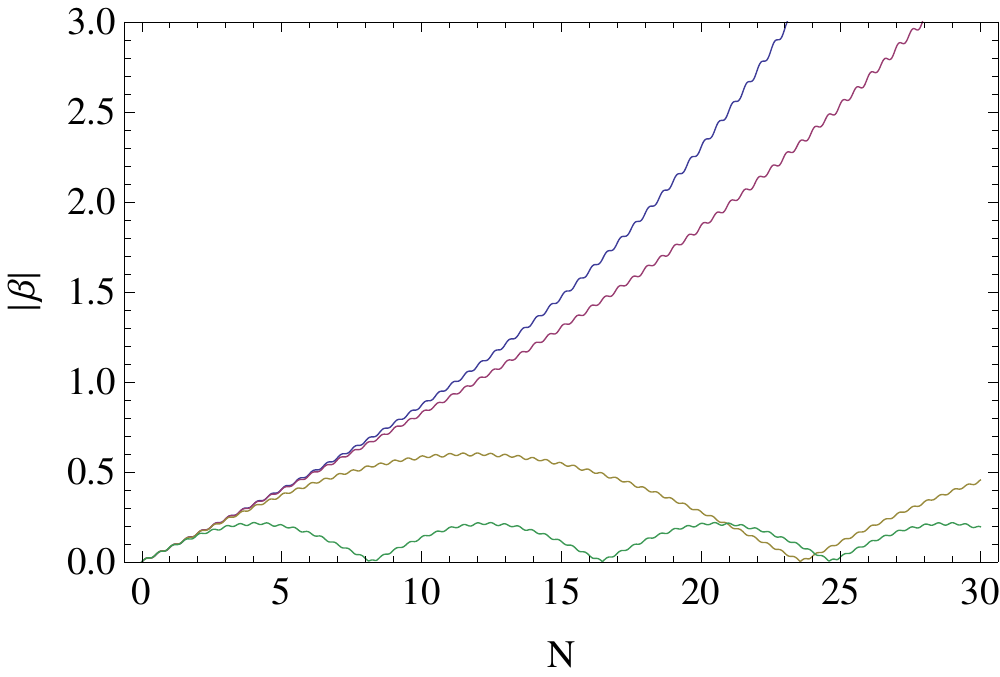}
\caption{Here is plotted $\abs{\beta}$ as a function of $N $ for various values of $R$, from top to bottom: $0$ (blue), $0.5$ (purple), $2$ (yellow), and $5$ (green). For all plots, the modulation amplitude $A=0.1$. After an initial linear growth for all curves, those with $R<1$ grow exponentially with $N$, while those with $R>1$ rise and fall periodically, the amplitude and period being approximately proportional to $1/R$. We also note the small rapid oscillations occurring on top of the long-time behavior.}
\label{fig:betaofN}
\end{minipage}
\newline \vspace*{3 mm} \newline
\begin{minipage}{1\linewidth}
\includegraphics[width=1\linewidth]{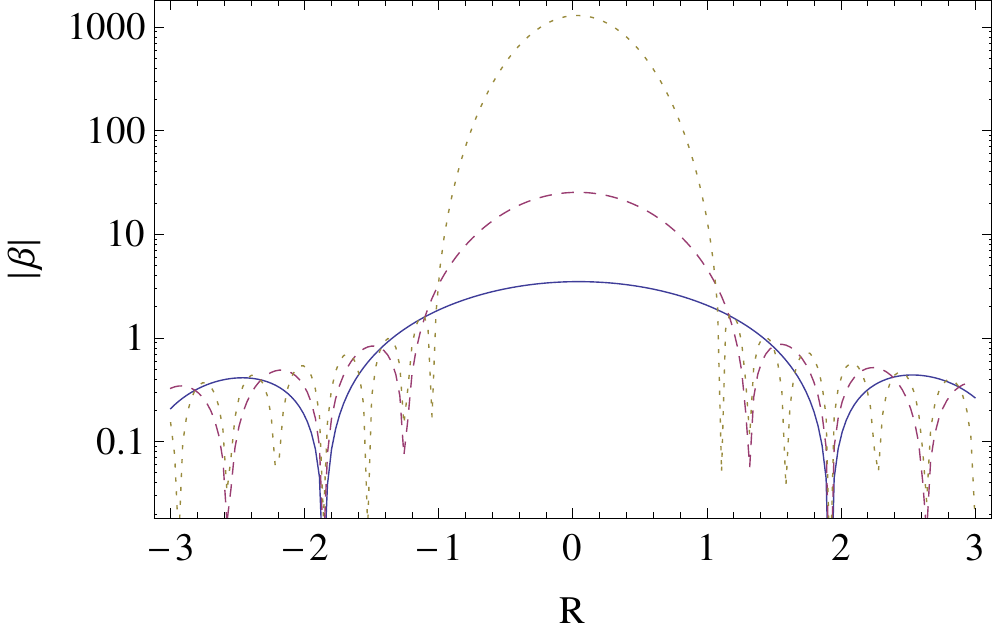}
\caption{Here is shown a logarithmic plot of $\abs{\beta}$ as a function of $R$ for various values of $N$: $25$ (blue solid), $50$ (purple dash), and $100$ (yellow dotted). For all plots, the modulation amplitude $A=0.1$. We clearly see the emergence of a central peak with increasing $N$, extending from $R=-1$ to $R=1$. For $\abs{R}>1$, the curves oscillate in a complicated way, as for small values of $\abs{\beta}$ the small rapid oscillations become more important. Because of these, $\abs{\beta}$ need not exactly vanish at the completion of a cycle. The number of long-time oscillations increases in proportion to $N$, and their maxima trace out an envelope corresponding to the $1/\sqrt{R^2-1}$ behavior of their maxima.}
\label{fig:betaofres}
\end{minipage}
\end{figure}

\begin{figure}[t]
\includegraphics[width=1\linewidth]{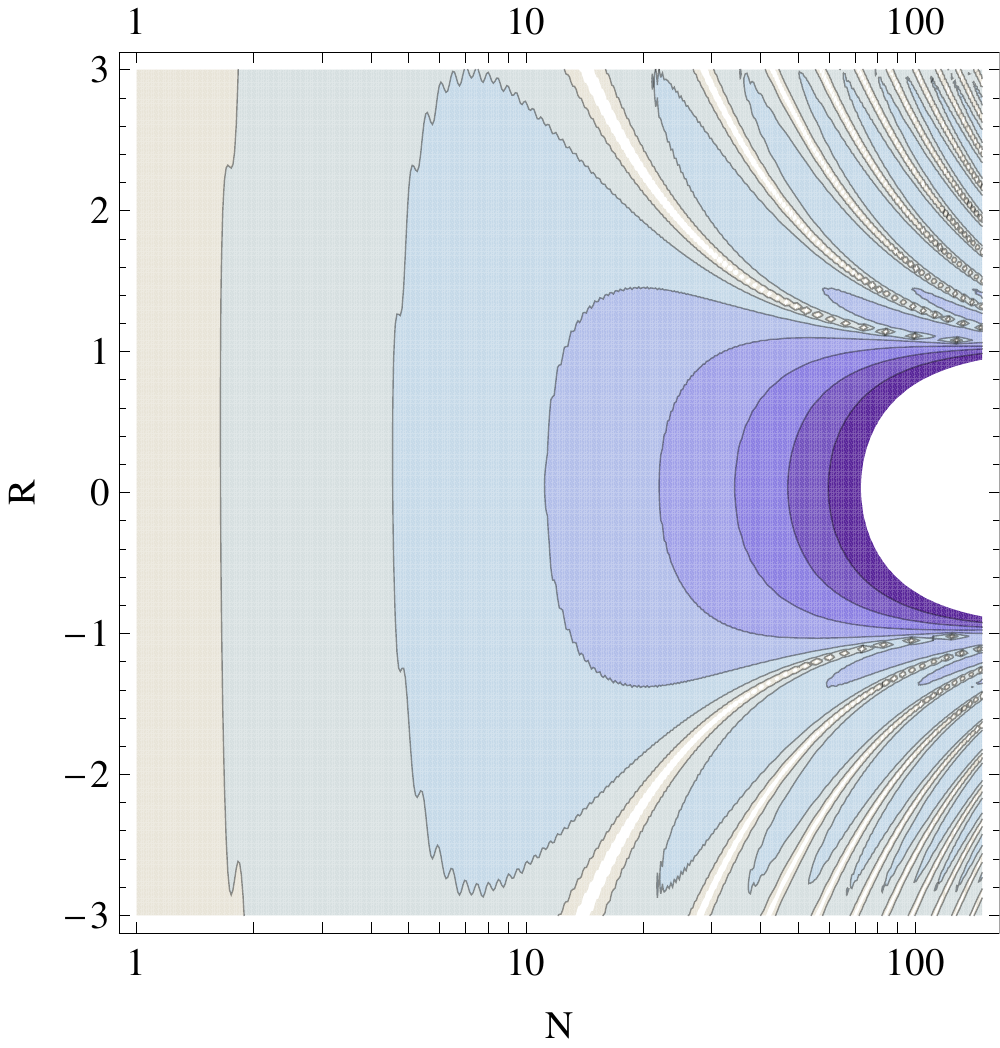}
\caption{Here is shown a contour plot of $\ln \abs{\beta}$ as a function of $N$ and $R$. As in Figs.~\ref{fig:betaofN} and~\ref{fig:betaofres}, the modulation amplitude $A=0.1$. The contour values are $5$ (the maximum shown, at the boundary between dark blue and white), $4$, and decrease by steps of $1$. We clearly see, with increasing $N$, the emergence of an exponentially growing resonant regime for $\abs{R}<1$ and an oscillating nonresonant regime for $\abs{R}>1$. Note also that the contours themselves are not exactly smooth, having jagged edges due to the short-time oscillations of $\abs{\beta}$.}
\label{fig:betaofresN}
\end{figure}

We consider an extended coherent modulation of the system that induces a corresponding modulation of $\omega_k^2$ of Eq.~\eqref{eq:omegak}. More precisely, $\omega_k^2$ is assumed to be constant for negative times, then follows a sinusoidal oscillation of duration $\Delta t$, and settles on a constant final value for all later times:
\begin{equation}
\omega_k^2(t)/\omega_0^2 = \left \{
\begin{array}{ll}
1 &\text{ if } t < 0 ,\\
1 + A \sin \omega_p t &\text{ if } 0 < t < \Delta t , \\
1 + A \sin \omega_p \Delta t &\text{ if } \Delta t < t .
\label{modul}
\end{array}\right .
\end{equation}
This function is illustrated in Fig.~\ref{fig:omegaoft}. It defines three dimensionless parameters that are all relevant in the following: the relative peak-to-peak\footnote{Assuming $A \ll 1$, the square root of the second line of Eq.~\eqref{modul} gives $\omega_k(t)/\omega_0 \approx 1 + \left(A/2\right) \sin \omega_p t$, so that the relative amplitude of the frequency modulation (as opposed to that of the {\it squared} frequency) is $A/2$.} amplitude $A$ of the frequency modulation, the number of oscillations $N$, and a resonance parameter we call $R$. Explicitly, these are defined by
\begin{equation}
\label{eq:3param}
\begin{split}
N &\doteq \omega_p \Delta t / 2\pi ,\\
AR/4 &\doteq \left( 2\omega_0 - \omega_p \right)/\omega_p. 
\end{split}
\end{equation}
Notice that $R$ combines in a particular way the amplitude $A$ and the detuning $\left( 2\omega_0 - \omega_p \right)/\omega_p$: it describes the relative frequency gap from resonance, scaled by $A$ so that it depends on this distance as a fraction of the ``width'' of the frequency modulation. Notice also that $N$ is not necessarily an integer.\footnote{
Recently, we became aware of Ref.~\cite{Westerberg:2014lra}, in which similar issues are considered in the context of nonlinear optics in thin films of material. They work at small $\abs{\beta}^2$, namely with the first term of Eq.~\eqref{eq:betaoffres2term}, and in the limit $A \to 0$. In this regime, the resonance occurs only at $\omega_p = 2 \omega_0$. Instead of working, as here, with an infinite homogeneous system over a finite time $\Delta t$, they work with a spatially finite system over an infinite time, and as a result the role of our temporal parameter $\Delta t$ is played by the side length of the film.}

\begin{figure*}
\begin{minipage}{0.47\linewidth}
\includegraphics[width=1\linewidth]{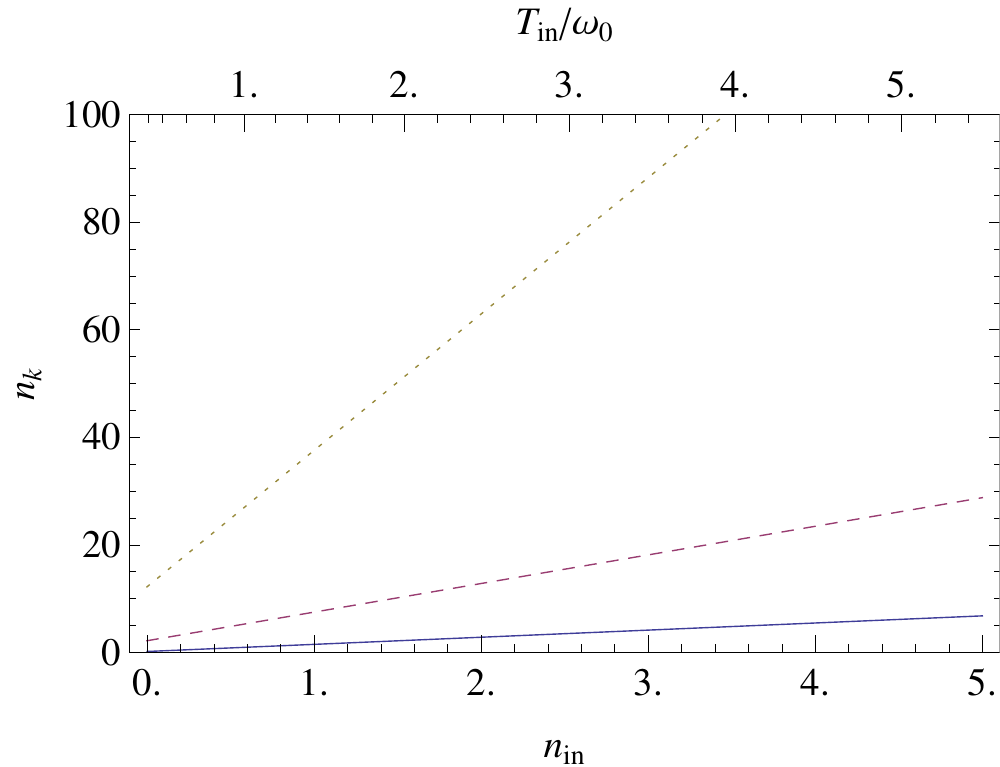}
\end{minipage}
\hspace{0.03\linewidth}
\begin{minipage}{0.47\linewidth}
\includegraphics[width=1\linewidth]{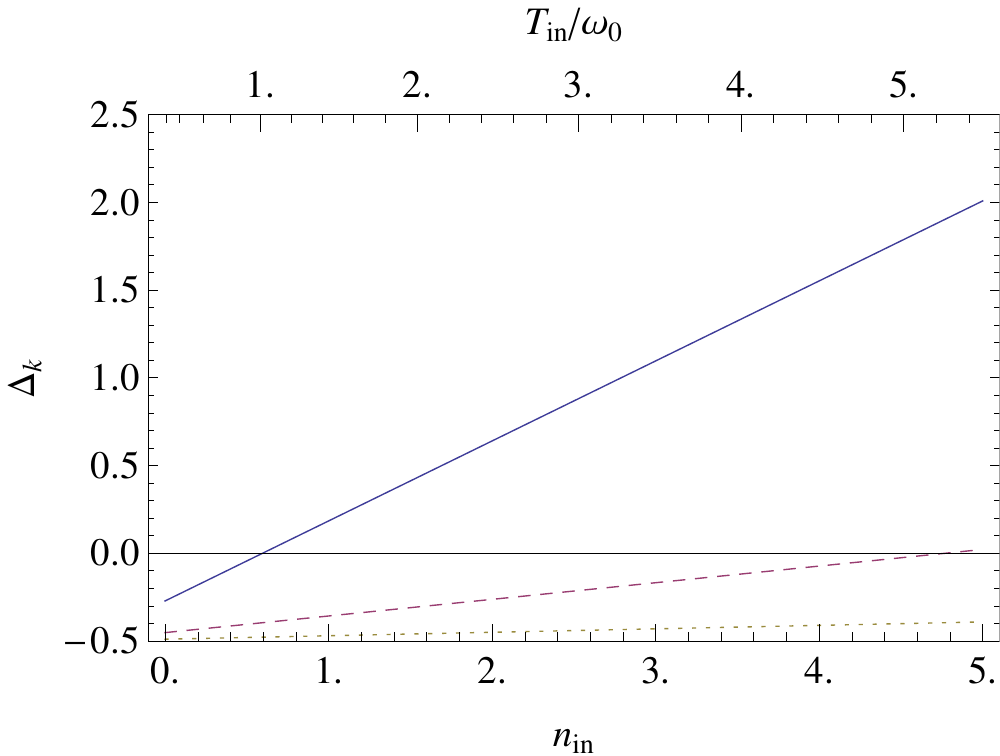}
\end{minipage}
\caption{Here are plotted, for exact resonance $R=0$, the final occupation number $n_{\rm out}$ (left figure) and $\Delta$ (right figure) as functions of the initial occupation number $n_{\rm in}$ (and the initial temperature in units of the mean frequency $\omega_0$) for various values of $N$: $5$ (blue solid), $15$ (dashed purple), and $25$ (dotted yellow). It is clear that both $n_{\rm out}$ and $\Delta$ increase with temperature in an approximately linear fashion. However, whereas increasing $N$ raises both the intercept and slope of the $n_{\rm out}$-curves, it has the opposite effect on the $\Delta$-curves, yielding a nonseparable state over a progressively wider range of initial temperatures.}
\label{fig:NDeltaofT}
\end{figure*}

For convenience, in the rest of the paper, we shall use $ N_t = \omega_p t / 2\pi $ as a dimensionless time parameter. Since $\alpha$ and $\beta$ are continuous in time, we can think of $\alpha(N)$ and $\beta(N)$ either as their final values after a modulation of length $N$, or as their instantaneous values at $N_t=N$ during a modulation of indeterminate length. These equivalent points of view allow us to use the same notation for $N$ and $N_t$, and also for $\abs{\beta^{\rm as}(N)}$ and $\abs{\beta(t = \Delta t_N)}$. 

\subsection{Properties of \texorpdfstring{\boldmath{$\abs{\beta}$}}{beta}} 

To reveal the various effects of the parametric amplification induced by Eq.~\eqref{modul}, we first numerically study the behavior of the norm of $\abs{\beta}$, whose square gives the final occupation number when working in the $in$ vacuum. In the body of the paper we provide only a qualitative description, but the following observations are all explained in the analytical study presented in Appendix~\ref{app:analytic}.

In Fig.~\ref{fig:betaofN}, we represent $\abs{\beta}$ as a function of adimensionalized time $N$, for various values of the parameter $R$. Each of the curves is a superposition of a large long-time variation, and a small short-time variation. The former is by far the most significant contribution, and shall be discussed below. The origin of the latter is provided in Eq.~\eqref{eq:betaoffres2term}. We notice that, after an initial linear growth whose rate is independent of $R$, the dependence of $\abs{\beta}$ on $N$ falls into one of two types, depending on the value of $R$:
\begin{enumerate}[i.]
\item For $\abs{R}>1$, $\abs{\beta}$ eventually drops below the linear curve, heading back towards zero and falling into a periodic oscillation. This is the off-resonant regime.
\item For $\abs{R}<1$, $\abs{\beta}$ eventually climbs above the linear curve, tending towards exponential growth. This is the resonant regime, and is due to stimulated amplification of formerly spontaneously created quanta.
\end{enumerate}

We thus get a sense from Fig.~\ref{fig:betaofN} of the importance of the parameter $R$. To investigate this further, in Fig.~\ref{fig:betaofres} we represent $\abs{\beta}$ in logarithmic scale as a function of $R$ for various values of $N$. We observe a central peak around $R=0$ of increasing height and decreasing width, though the width saturates with increasing $N$ such that it extends from $R=-1$ to $1$. This peak corresponds to the resonant regime, and as the evolution of $\abs{\beta}$ becomes exponential for large $N$ we find that the height of the peak in $\log \abs{\beta}$ varies linearly with $N$. For $\abs{R}>1$, we observe oscillations with a fixed maximum value for each value of $R$, such that the maxima trace out an envelope that is a smooth function of $R$. This is the nonresonant regime, where $\abs{\beta}$ never rises above a fixed maximum value.

For completeness, in Fig.~\ref{fig:betaofresN} we combine the above figures in a contour plot, where the contours are lines of constant $\abs{\beta}$ in the $(N,R)$-plane. One can clearly see the emergence of the resonant peak for $\abs{R}<1$ with increasing $N$, as well as the oscillatory behavior for $\abs{R}>1$.

\begin{figure}[htb]
\includegraphics[width=1\linewidth]{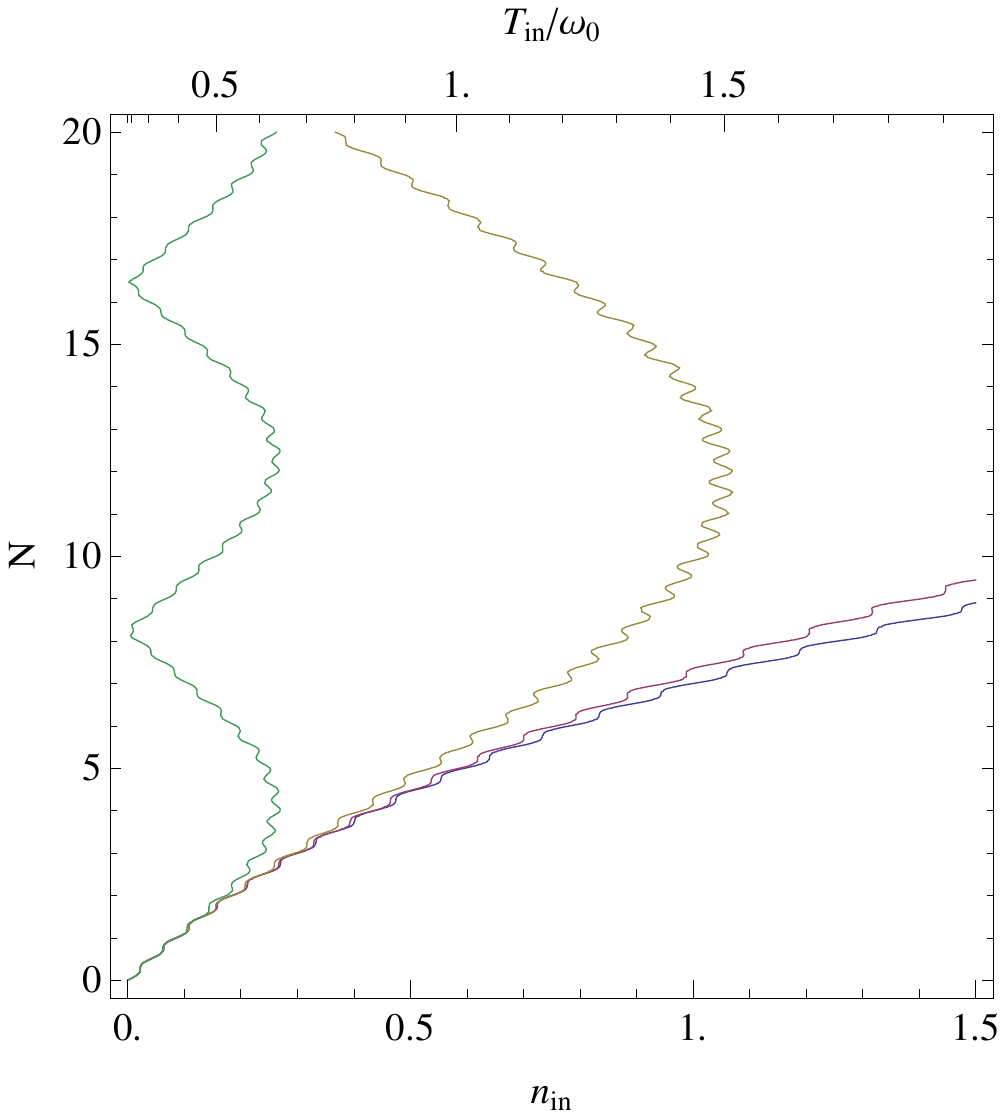}
\caption{Plotted here are loci of the separability threshold $\Delta=0$ in the $(T_{\rm in},N)$-plane for various values of $R$, from right to left: $0$ (blue), $0.99$ (purple), $2$ (yellow), and $5$ (green). As for previous plots, we work with $A=0.1$. Note once again the splitting of the large-$N$ behavior into resonant ($\abs{R}<1$) and nonresonant ($\abs{R}>1$) regimes. For $R<1$, the curves are seen to extend indefinitely towards higher values of $n_{\rm in}$, meaning that it is possible, for any initial temperature, to reach a nonseparable state if $N$ is made large enough. On the other hand, for $R>1$, the curves reach some maximum value of $n_{\rm in}$ and then turn around, so the state can only be made nonseparable for temperatures below some maximum value, and even then the system will oscillate between separable and nonseparable states. Again we notice the lack of smoothness of the curves due to the short-time oscillations of $\abs{\beta}$.}
\label{fig:DeltaofNT}
\end{figure}

\subsection{Dependence on temperature and final entanglement}

We now include the effects of a nonzero initial temperature -- or, equivalently, a non-zero initial occupation number -- on both the final occupation number $n_{\rm out}$ and the separability parameter $\Delta_{\rm out}$.  Through Eqs.~\eqref{eq:outnc}-\eqref{eq:deltaoutbeta}, the latter quantities are directly related to the initial occupation number, which is in turn related to the initial temperature via the Planck distribution
\begin{equation}
\label{eq:thermal}
n_{th}(\omega/T) = \frac{1}{e^{\omega/T}-1} .
\end{equation}
In Fig.~\ref{fig:NDeltaofT}, $n_{\rm out}$ and $\Delta_{\rm out}$ are plotted as functions of $n_{\rm in}$ and $T/\omega_0$ for a system exactly at resonance ($R=0$) and for various values of $N$. We observe that $n_{\rm out}$ increases both with initial temperature and with $N$, while $\Delta$ -- which is sensitive to the division of $n_{\rm out}$ into spontaneous and stimulated contributions -- increases with initial temperature but \textit{decreases} with $N$. This is in accordance with Eq.~\eqref{eq:deltaoutbeta} since $\abs{\beta}$ increases with the duration $N$.

In Fig.~\ref{fig:DeltaofNT}, we represent the nonseparability threshold in the $(N,T)$-plane -- that is, the locus where $\Delta=0$ -- for various values of $R$. Notice that $\Delta$ is positive to the right of the curves since it always increases with $n_{\rm in}$. In the case of resonance ($\abs{R}<1$), we observe that whatever the initial temperature, $\Delta$ becomes negative and the state becomes nonseparable for $N$ larger than some value. By contrast, in the nonresonant case ($\abs{R}>1$), there exists a temperature above which the state is separable for all values of $N$. This critical temperature depends on $R$ and is generically lower than $\omega_0$. The analytic treatment of Appendix~\ref{app:analytic} gives $T_{\max} \sim \omega_0/ \ln(R)$.

To conclude this section, we consider the experiment of Ref.~\cite{PhysRevLett.109.220401}, the results of which triggered the present analysis. From the data, we estimate that the duration $N \sim 50$, and that the peak-to-peak amplitude $A \sim 0.1$. (In fact, this is only an upper bound on $A$. It is actually the frequency of the trapping potential that is modulated with this amplitude, and estimating the corresponding amplitude for the mode frequencies is rather nontrivial.) We have not been able to determine with precision the appropriate value for $R$. In principle, if one works exactly at the resonance $R = 0$, the phonon state would be nonseparable for $n_{\rm in} \lesssim 1200 $. Instead, when working with $R = 1$, nonseparability would occur only for $n_{\rm in} \lesssim 40$. These findings seem to overestimate the observed intensity of the correlations. A possible explanation for this is the neglect of weak dissipative effects. To study this possibility, in Sec.~\ref{sec:dissip} we include weak dissipation while the system is being modulated. We shall see that weak dissipation is sufficient to ruin the nonseparability reached by the system in the absence of dissipation, thereby possibly explaining what was reported in Ref.~\cite{PhysRevLett.109.220401}. 

\section{Weak dissipation} 
\label{sec:dissip}

In this section, we introduce a weak dissipative rate $\Gamma$ -- where ``weak'' means $\Gamma/\omega_0 \ll 1$ -- and study its effects on quasiparticle creation and entanglement.

\subsection{Dissipation model}

In the presence of dissipation, the notion of Bogoliubov coefficients is no longer well defined, as the system is coupled to some environment. As a result, the state of the system can no longer be characterized by $\alpha(t)$ and $\beta(t)$ as in the non-dissipative case. Instead, the mean occupation number $n$ and the correlation $c$ can still be defined for all times when dissipation is weak enough~\cite{Busch:2013gna,Busch:2013sma}. In this regime, the separability parameter $\Delta$ remains related to these by Eq.~\eqref{Ineq}. 

Our model of dissipation is phenomenological and based on the results of Refs.~\cite{Busch:2013gna,Busch:2013sma}.  Its details and the derivation of the equations of motion are given in  Appendix~\ref{app:dissip}.  There it is found that, assuming the environment to be incoherent, $n$ and $c$ evolve according to the following first order coupled equations: 
\begin{equation}
\label{eq:eomdissipncexact}
\begin{split}
(\partial_t+ 2 \Gamma)n &= 2 \Gamma n_{eq} + \frac{\dot u}{v} \Re \left [ \ep{-2 i \int \omega dt} c \right ] , \\
(\partial_t+ 2 \Gamma)c &= \frac{\dot u}{v} \ep{2 i \int \omega dt} (1+2n) .
\end{split}
\end{equation}
Here, $n_{eq}(t) = n_{eq}(\omega_k(t))$, where $n_{eq}(\omega)$ is the mean occupation number at frequency $\omega$ when the system reaches equilibrium in the limit $t \rightarrow \infty$. It is determined by the state of the environment, and is typically given by $n_{eq}(\omega) = n_{th}(\omega/T)$, for given temperature $T$ and where $n_{th}(\omega/T)$ is the thermal distribution of Eq.~\eqref{eq:thermal}.

\begin{figure*}
\includegraphics[width=0.45\linewidth]{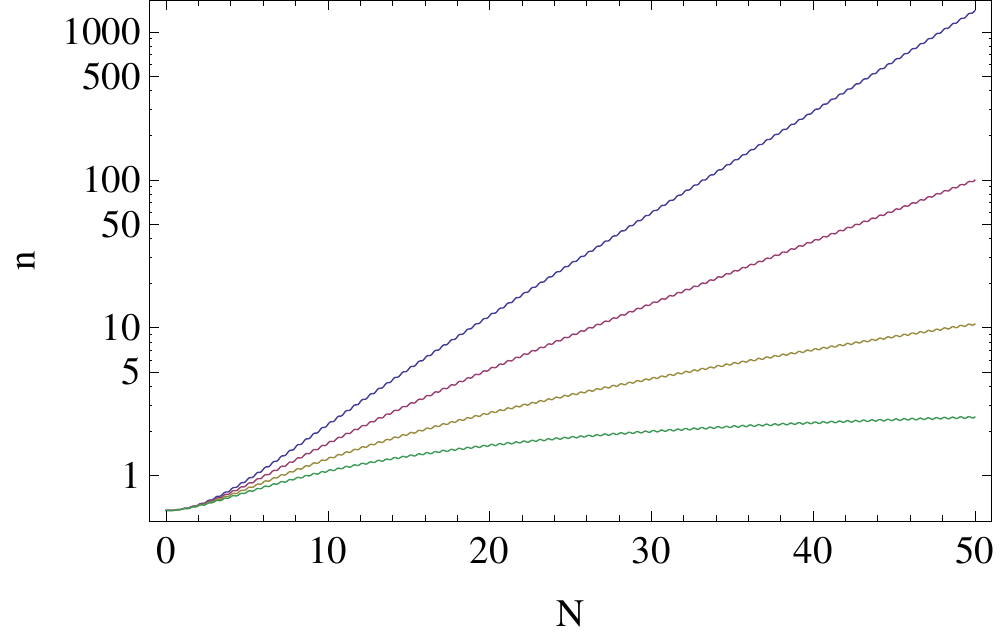}
\hspace{0.03\linewidth}
\includegraphics[width=0.45\linewidth]{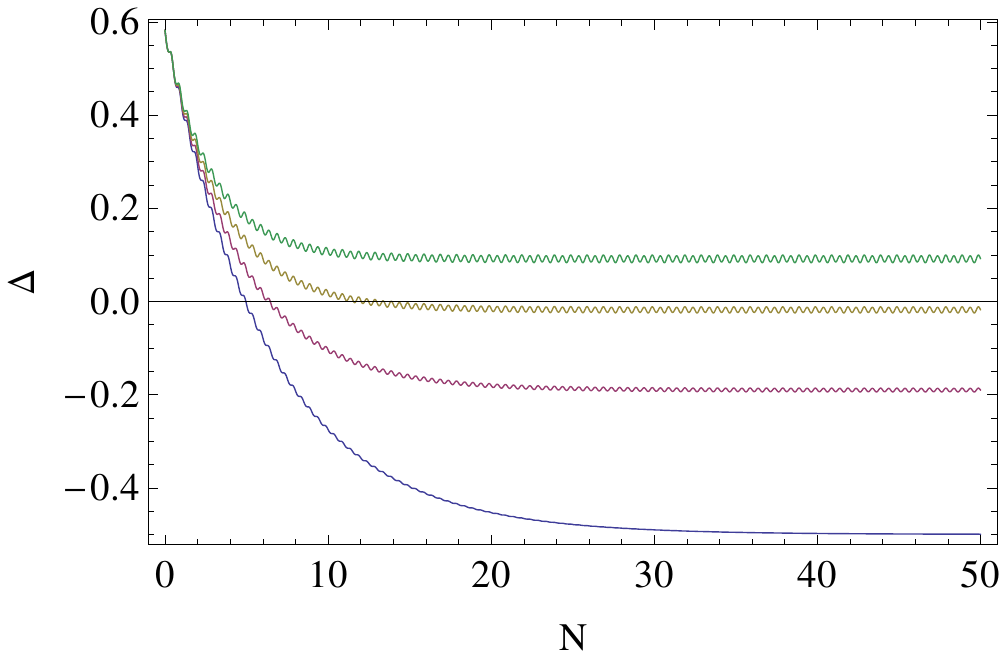}
\caption{Here are plotted, at exact resonance $R=0$ and for an initial thermal bath with $T_{\rm in}=\omega_0$, the mean occupation number (left figure) and the separability parameter $\Delta$ (right figure) immediately after the end of the frequency modulation. The amplitude is as before taken to be $A=0.1$. The various curves correspond to different dissipative rates $\Gamma/\omega_0$ (from larger to smaller $n$ and from lower to higher $\Delta$): $0$ (blue), $0.01$ (purple), $0.02$ (yellow) and $0.03$ (green). The rate of increase of $n$ is seen to be reduced by larger dissipation rates; but, in accordance with the prediction of Sec.~\ref{sec:dissip}, it approaches exponential growth for $\Gamma/\omega_0 < A/4 = 0.025$ and saturates for $\Gamma/\omega_0>0.025$. Similarly, $\Delta$ approaches a limiting value which increases with the dissipation rate, and as predicted in Sec.~\ref{sec:dissip} the final state is nonseparable only when $\Gamma/\omega_0 < A/8n_{eq} \approx 0.021$.}
\label{fig:ndissiphighT}
\end{figure*}

\begin{figure*}
\includegraphics[width=0.45\linewidth]{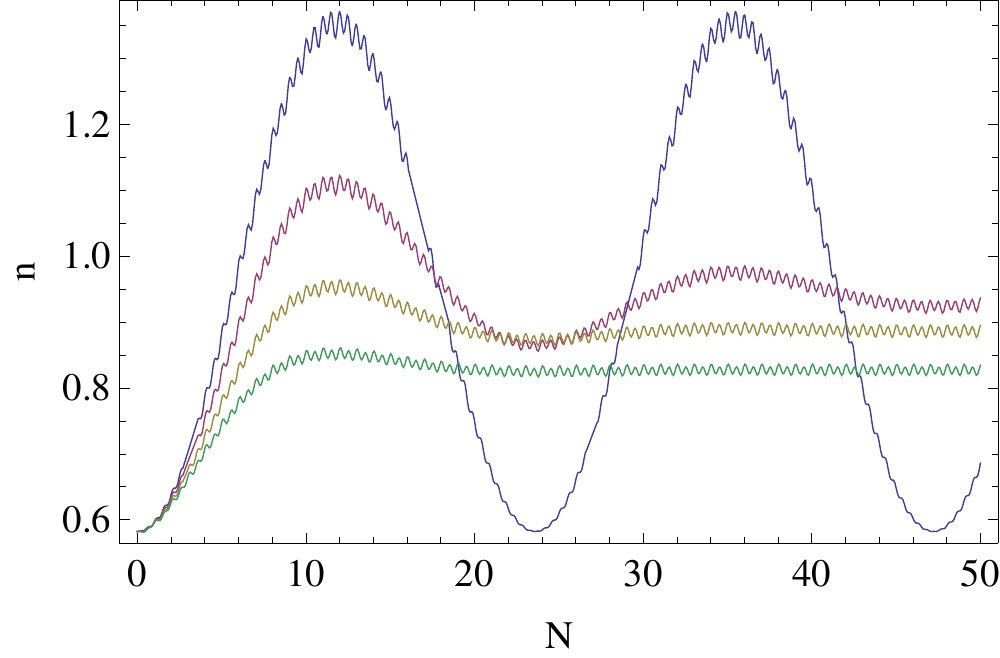}
\hspace{0.03\linewidth}
\includegraphics[width=0.45\linewidth]{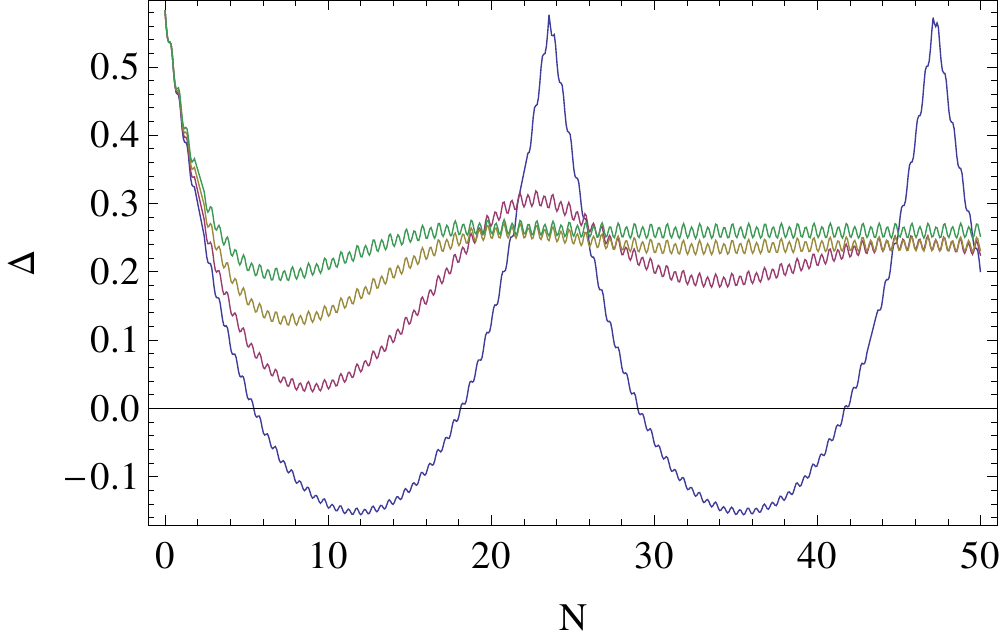}
\caption{Here are shown, at $R=2$ and for an initial thermal bath with $T=\omega_0$, the mean occupation number (left figure) and separability parameter $\Delta$ (right figure) immediately after the end of the frequency modulation. As before, the amplitude $A=0.1$. The various curves correspond to different values of the dissipative rate $\Gamma/\omega_0$, from larger to smaller first oscillation: $0$ (blue), $0.01$ (purple), $0.02$ (yellow) and $0.03$ (green). We note a smoothing out of the oscillations with increasing dissipative rate, and eventually their disappearance, as in overdamped systems. $n$ and $\Delta$ are seen to approach limiting values, which (respectively) decrease and slightly increase with increasing $\Gamma/\omega_0$.}
\label{fig:ndissipoffres}
\end{figure*}

We work in a regime where the relative modulation of the mode frequency is small, and where it is not too far from resonance; in the Appendixes, we shall see that this regime corresponds to $A\ll 1 $, $AR/4 \ll 1 $.  Thus we can approximate $ \frac{\dot u}{v} \ep{-2 i \int \omega dt} \sim \frac{A}{4} \omega_p \ep{- i AR \omega_p t/4} $, and
if we also average over the rapid oscillations so that these can be neglected, Eqs.~\eqref{eq:eomdissipncexact} become 
\begin{equation}
\label{eq:eomncdissipapprox}
\begin{split}
(\partial_t+ 2 \Gamma)n &= 2 \Gamma n_{eq} + \frac{A}{4} \omega_p \Re \left[ \ep{-i AR \omega_p t/4} c \right] ,\\
(\partial_t+ 2 \Gamma)c &=\frac{\omega_p A}{8} \ep{i AR\omega_p t /4 } (1+2n) . 
\end{split}
\end{equation}
Neglecting the time dependence of $n_{eq}$ and working at exact resonance $R=0$, it can be derived from Eqs.~\eqref{eq:eomncdissipapprox} that, for $\Gamma/\omega_0>A/4$, $n$ grows and saturates at
\begin{equation}
\label{eq:nsaturate}
\begin{split}
n_{max} = \frac{32 \Gamma^2 n_{eq} + A^2 \omega_0^2}{32 \Gamma^2 - 2 A^2 \omega_0^2}.
\end{split}
\end{equation}
On the other hand, for $\Gamma/\omega_0 < A/4$, $n$ grows exponentially, albeit at a slower rate due to dissipation. In both these cases, $\Delta$ decays exponentially towards the limiting value
\begin{equation}
\label{eq:Deltasaturate}
\begin{split}
\Delta_{min} = \frac{8 \Gamma n_{eq} - A \omega_0}{8 \Gamma + 2 A \omega_0}.
\end{split}
\end{equation}
Thus, the state eventually becomes nonseparable if $2n_{eq} \Gamma/\omega_0 < A/4$. Note that this condition for nonseparability is independent of the condition for exponential growth of $n$.  More precisely, if $n_{eq}<1/2$, there exists a regime where $n$ saturates and the final state is nonseparable, whereas if $n_{eq}>1/2$, there exists a regime in which the final state is separable and $n$ grows exponentially. 

\subsection{Numerical analysis}

For the results presented here, we take for initial and equilibrium values of $n$ the thermal distribution of Eq.~\eqref{eq:thermal} and for equation of motion Eq.~\eqref{eq:eomdissipncexact}. In Fig.~\ref{fig:ndissiphighT} we represent, for various dissipative rates $\Gamma/\omega_0$, the time evolution of $n$ and $\Delta$ at exact resonance $R=0$ and with a temperature for the environment $T=\omega_0$. We observe that the mean occupation number decreases with $\Gamma/\omega_0$, the deviation becoming larger with increasing $N$; and that, for large enough dissipative rates, $n$ is seen to saturate at large $N$. Correspondingly, the coherence is reduced (i.e., $\Delta$ increases), and for large enough $\Gamma/\omega_0$ the limiting value of $\Delta$ is positive so that the state never becomes nonseparable. As an example, with the numbers of Ref.~\cite{PhysRevLett.109.220401}, 
namely for $T_{\rm in}/ \omega_0 = 1$ and $N = 50 $, a weak dissipation of $ \Gamma / \omega_0 =2\% $ is almost sufficient to ruin the nonseparability which is found in the absence of dissipation. To be more explicit, in the absence of dissipation, $\Delta =- 0.4995 $, while for $\Gamma / \omega_0 =2\%$, $\Delta = -0.018 $. Nonseparability is lost for $\Gamma / \omega_0 \sim 2.1\%$. 

Slightly off resonance, with $0 < \abs{R} < 1$, the differences with respect to the resonant case are similar to those of the non dissipative case.  See Fig.~\ref{fig:betaofres} for the behavior of the mean occupation number $n_{out}=\abs{\beta}^2$ and Fig.~\ref{fig:DeltaofNT} for the behavior of the separability parameter $\Delta$ for $0<\abs{R}<1$ in the absence of dissipation.  As can be seen there, $n_{out}$ falls with $R$ and $\Delta$ increases with $R$.

In the nonresonant case, with $R>1$, dissipation is observed to dampen the oscillations in $n$ and $\Delta$, both of which approach limiting values that (respectively) decrease and increase with increasing $\Gamma/\omega_0$; see Fig.~\ref{fig:ndissipoffres}. It is even possible to reach an overdamped regime where no oscillations occur. As in the absence of dissipation, on top of this long range behavior some small and rapid oscillations of frequency near $2\omega_p$ occur. These do not decay when the system reaches a near-equilibrium state, as can be verified by considering the near-stationary solution of Eqs.~\eqref{eq:eomncdissipapprox} when the rapid oscillations are taken into account. 

We conclude this section by applying our results to the experiment described in Ref.~\cite{Lahteenmaki12022013}. We find that the relevant parameters are $n_{eq}=0.0056$, $A \approx 0.048$ and $\Gamma/\omega_0 > 0.009$. (We can only give a lower bound for $\Gamma/\omega_0$ because it is acknowledged that there is additional source of dissipation -- probably two-photon dissipation -- that is not accounted for.) Assuming the experiment is performed very close to resonance, we take $R=0$, so Eqs.~\eqref{eq:nsaturate} and~\eqref{eq:Deltasaturate} are applicable. Since $A/8n_{eq} \approx 1 > \Gamma/\omega_0$, we conclude that $-1/4 \leq \Delta < 0$ and the state is nonseparable. This is in agreement with the results of Ref.~\cite{Lahteenmaki12022013} which reports $\Delta = \left(2^{-0.32}-1\right)/2 \approx -0.1$. The behavior of $n$, however, is more difficult to ascertain since $A/4$ is slightly above the lower bound of $\Gamma/\omega_0$.  We expect that the additional dissipative effects will take $\Gamma/\omega_0$ above $A/4$, so that $n$ should saturate. 

\section{Conclusions}
\label{sec:conclusion}

We have considered the spectrum of quasiparticles and their degree of entanglement due to a sinusoidal modulation of the (squared) frequency in a homogeneous quantum system. For definiteness, the system under consideration was taken to be an atomic Bose-Einstein condensate. The modulation was found to be describable by three parameters: its length, its amplitude, and the detuning of its frequency from resonance (at twice the mean mode frequency). The final amount of spontaneous creation, described by the magnitude of the Bogoliubov coefficient $\abs{\beta}$, is found to have a complicated dependence on these three parameters, with the behavior of the separability parameter $\Delta_{\rm out}$ following suit via Eq.~\eqref{eq:deltaoutbeta}.

A key observation, in accordance with similar results seen in Ref.~\cite{Kofman:1997yn}, is the existence of a finite width of ``resonant'' frequencies. Averaging out the small rapid oscillations that are superimposed on a large long-time behavior, the spontaneous contribution to resonant quasiparticle modes grows exponentially with the duration of the modulation, and for any initial temperature, the final state can be made nonseparable if the modulation lasts long enough. For off-resonant modes, however, the spontaneous contribution rises and falls periodically, never reaching above some maximum value. At the level of entanglement, this has the effect that, for off-resonant modes, there is a temperature above which the final state is always separable, no matter the length of the modulation.

Finally, we evaluated the consequences of weakly dissipative effects. We demonstrated that the nonseparability of the final state can be significantly reduced and even destroyed when these are taken into account. It is thus clear that weak dissipation could play an important role in the experimental attempts to establish nonseparability of the final state. These considerations have been illustrated by considering two recent experiments. 

\acknowledgments
We are grateful to Chris Westbrook and Iacopo Carusotto for discussions during the first QEAGE workshop, and for interesting comments on the first version of this work. This work was supported by the French National Research Agency under the Program Investing in the Future Grant No.~ANR-11-IDEX-0003-02 associated with the project QEAGE (Quantum Effects in Analogue Gravity Experiments).

\appendix

\section{Analytical properties of \texorpdfstring{$\boldsymbol{\abs{\beta}}$}{modulus of beta}} 
\label{app:analytic}

As can be seen from Figs~\ref{fig:betaofN} and~\ref{fig:betaofres}, the dependence of $\abs{\beta}$ on the parameters $N$, $A$ and $R$ is rather complicated. Yet, the essential features can be obtained analytically, as we now show.

To simplify the following equations, we use the adimensional time $\tau$ and the detuning parameter $r$ given by
\begin{equation}
\label{eq:tau-r}
\begin{split}
\tau \doteq \omega_p t, 
\quad r \doteq \frac{ 2\omega_0 - \omega_p}{\omega_p} = AR/4 .
\end{split}
\end{equation}
We also assume that $A \ll 1$, so the relative modulation of $\omega$ is small. Then Eqs.~\eqref{eq:eomalphabeta} simplify and become
\begin{subequations}
\label{eq:eomalphabetaapprox1}
\begin{align}
\label{eq:eomapproxalpha}
\partial_{\tau } \alpha &\approx \frac{A}{8} \left[ e^{i r \tau } + e^{i(2- r)\tau } \right] \beta ,\\
\label{eq:eomapproxbeta}
\partial_{\tau } \beta &\approx \frac{A}{8} \left[ e^{-i r\tau } + e^{-i(2-r)\tau } \right] \alpha .
\end{align}
\end{subequations}
To solve these equations, two cases will be separately considered: in the first, the modulation is nonresonant so $\beta \ll 1$ for all times; in the second, the modulation is close to resonant so $A R /4 = r \ll 1$.

\subsection{Non-resonant case}

When $\beta$ is very small, unitarity $\abs{\alpha}^2 =1+ \abs{\beta}^2 $ implies that $\abs{\alpha}$ remains close to $1$. Equation~\eqref{eq:eomapproxalpha} then guarantees that the phase of $\alpha$ is slowly varying in time, so $\partial_t(\beta / \alpha) \sim \partial_t(\beta) / \alpha$. Since we seek only the magnitude $\abs{\beta}$, we shall not consider this phase. We thus have
\begin{equation}
\label{eq:dtbetaoffres}
\begin{split}
\partial_{\tau } \beta &\approx \frac{A}{8} \left[ e^{-i r\tau } + e^{-i(2-r)\tau } \right] .
\end{split}
\end{equation}
This is trivially solved by
\begin{equation}
\label{eq:betaoffres2term}
\begin{split}
\beta(t) &\approx \frac{-A}{8} \left [ \frac{e^{-i \tau r}-1}{ r} + \frac{e^{-i \tau (2+r)}-1}{2 +r} \right ].
\end{split}
\end{equation}
This equation correctly describes two effects that are visible in Figs.~\ref{fig:betaofN} and~\ref{fig:betaofres}: The first term describes long-time variations of large magnitude, while the second describes short-time variations of small magnitude.

\subsection{Close to resonance}

We now suppose that we are close to resonance so $r \ll 1$. In such a case, a rotating wave approximation can be performed so that we neglect terms oscillating with frequency $2 \omega_0 + \omega_p$. Under such circumstances, the Bogoliubov coefficients are solutions of 
\begin{equation}
\label{eq:eomalphabetaapprox}
\begin{split}
\partial_{\tau } \alpha &\approx \frac{A}{8} e^{i r \tau } \beta ,\\
\partial_{\tau } \beta &\approx \frac{A}{8} e^{-i r \tau } \alpha .
\end{split}
\end{equation}
Imposing the initial conditions $\alpha=1$ and $\beta=0$ at $t=0$, the exact solutions of these equations are
\begin{subequations}
\label{eq:alphabetasinh}
\begin{align}
\begin{split}
\alpha(t) \sim \ep{i \frac{r}{2} \tau } &\Bigg [ \cosh \left [ \frac{A}8 \sqrt{1-R^2}\tau \right ] \\
 &- i R\frac{\sinh \left [ \frac{A}8 \sqrt{1-R^2}\tau \right ]}{\sqrt{1-R^2}} \Bigg ] ,
 \end{split} \\
\beta(t) \sim \ep{-i \frac{r}{2} \tau } &\frac{\sinh \left [ \frac{A}8 \sqrt{1-R^2}\tau \right ]}{\sqrt{1-R^2}}.
 \label{eq:betasinh}
\end{align}
\end{subequations} 
Several comments should be made. First, for low values of $\tau =\omega_p t$, we have $\abs{\beta} \propto A \tau $. This explains the fact that all curves of Fig.~\ref{fig:betaofN} are initially linear with a growth rate that is independent of $R$. 

Second, Eqs.~\eqref{eq:alphabetasinh} reveal the crucial role played by $R$, which did not appear in Eqs.~\eqref{eq:eomapproxalpha}. The value of $R$ delineates the two behaviors that we observed, and characterizes the transition from one to the other occurring at $\abs{R}=1$. When $\abs{R}>1$, the square root is imaginary and $\beta$ oscillates in time, with a maximum given by $\abs{\beta}_{\rm max} \sim 1/\sqrt{R^2-1}$. This is the off-resonant behavior. In addition, the fact that $\abs{\beta}_{\rm max}$ depends only on $R$ explains the envelope traced out with increasing $N$ in Fig.~\ref{fig:betaofres}. In contrast, when $\abs{R} < 1$, $\beta$ grows exponentially, as can be clearly seen in the low-$R$ curves of Fig.~\ref{fig:betaofN}. This is the resonant regime, and the fact that it occurs over a finite range of $R$ explains the finite width of the growing part of the spectrum seen in Figs.~\ref{fig:betaofres} and~\ref{fig:betaofresN}. Indeed, at large times, we find $\ln \abs{\beta} \sim (\pi NA/4) \sqrt{1-R^2} - \ln(2 \sqrt{1-R^2})$. The critical case is $\abs{R}=1$. In this case, under the assumptions we used, $\beta$ grows linearly in time.

Third, in the limit $R \gg 1$, $r = A R/4 \ll 1$, Eq.~\eqref{eq:betasinh} gives
\begin{equation}
\begin{split}
\abs{\beta} \sim \frac{\sin \left [ r \tau /2 \right ]}{ R }, 
\end{split}
\end{equation}
which corresponds to the first term of Eq.~\eqref{eq:betaoffres2term}. There is thus a perfect compatibility of the two descriptions in this intermediate range $1 \ll R \ll 4/A$ where they overlap. 

Finally, we can substitute the expressions of Eq.~\eqref{eq:alphabetasinh} into the right-hand side of Eqs.~\eqref{eq:eomalphabetaapprox1}, yielding improved solutions that are relevant close to resonance and include the rapidly oscillating terms. In fact, iterating this operation gives a perturbative expansion for the solutions of Eqs.~\eqref{eq:eomalphabetaapprox1}, of which Eqs.~\eqref{eq:alphabetasinh} are the lowest-order terms.

\section{Model of weak dissipation}
\label{app:dissip}

We adopt a simple effective approach to dissipation, inspired by the results of Refs.~\cite{Busch:2013gna,Busch:2013sma} in which it was incorporated using Hamiltonian models that respect unitarity. In these references, only single sudden changes were considered, and it was found that the Bogoliubov coefficients -- which can be \textit{locally} defined in the vicinity of the change when dissipation is weak enough -- respond to the sudden change as if dissipation were not present. In fact, the main effect of weak dissipation observed was the expected exponential damping of the system towards an equilibrium state. 

These observations are here implemented by considering a series of infinitesimal double steps of duration $dt$. In each double step, the system evolves according to two processes:
\begin{enumerate}[i.]
\item a non-dissipative modulation linking $[n(t),c(t)]$ to an intermediate $[\tilde n(t), \tilde c(t)]$ by an infinitesimal Bogoliubov transformation $\delta S$, which can be derived from the local behavior of $\alpha(t)$ and $\beta(t)$ in the absence of dissipation; 
\item an exponential damping due to dissipation which carries $[\tilde n(t), \tilde c(t)]$ to $[ n(t+dt), c(t+dt)]$.
\end{enumerate}

The non-dissipative modulation over $dt$ gives rise to a small change in the quantum amplitude operators as in Eq.~\eqref{eq:bogoliubov}:
\begin{equation}
\label{eq:infoperators}
\hat{b}_{\bk}(t+dt) = \delta\alpha \, \hat{b}_{\bk}(t) + \left(\delta\beta\right)^{*} \hat{b}^{\dagger}_{-\bk}(t) ,
\end{equation}
where $\delta\alpha-1$ and $\delta\beta$ are proportional to $dt$. As described above, this is the same infinitesimal transformation that acts locally in the absence of dissipation, so if $S(t)$ is the finite Bogoliubov transformation relating the instantaneous amplitude operators $\hat{b}_{\bk}(t)$ and $\hat{b}^{\dagger}_{-\bk}(t)$ to $\hat{b}^{in}_{\bk}$ and $\left(\hat{b}^{in}_{-\bk}\right)^{\dagger}$ in the non-dissipative case, then $\delta S = S(t+dt) S^{-1}(t)$. In terms of $\alpha(t)$ and $\beta(t)$, this is equivalent to 
\begin{equation}
\label{eq:infbog}
\begin{split}
\delta \alpha &= \alpha(t+dt) \alpha^*(t) - \beta(t+dt)^* \beta(t) \\
&\sim 1+ (\dot\alpha \alpha^* - \dot\beta^* \beta)dt  , \\
\delta \beta &= \beta(t+dt) \alpha(t)^* - \alpha(t+dt)^* \beta(t) \\
&\sim ( \dot\beta \alpha^* -\dot\alpha^* \beta)dt .
\end{split}
\end{equation}
Moreover, unitarity requires that $\abs{\delta\alpha}^{2}-\abs{\delta\beta}^{2}=1$, so the difference $\abs{\delta\alpha}^{2} - 1$ is second-order in $dt$. Thus we are led to the following equations for the resulting changes in $n$ and $c$, see Eq.~\eqref{eq:outnc}:
\begin{equation}
\label{eq:infmod}
\begin{split}
\tilde n(t) &= \abs{\delta \beta}^2 + \left ( \abs{\delta \beta}^2+ \abs{\delta \alpha}^2 \right ) n(t) + 2 \Re\left (\delta \alpha \delta \beta c(t) \right )\\
&\sim n(t) + 2 \Re\left [\delta \beta c(t) \right ] , \\
\tilde c(t) &= \delta \alpha \delta \beta^{*} + 2 \delta \alpha \delta \beta^{*} n(t) + \delta \alpha^2 c(t) + \left(\delta \beta^{*}\right)^2 c^{*}(t)  \\
&\sim \delta \beta^{*}[ 1 + 2 n(t) ] + \delta \alpha^2 c(t) .
\end{split}
\end{equation}
In the second part of the double step, we account for the process of weak dissipation, which is described by
\begin{equation}
\label{eq:infdiss}
\begin{split}
n(t+dt) &= n_{eq}(t) + [\tilde n(t)-n_{eq}(t)] \ep{- 2 \Gamma dt}\\
&\sim \tilde n(t) - 2 \Gamma [\tilde n(t)-n_{eq}(t) ] dt , \\
c(t+dt) &= \tilde c(t) \ep{- 2 \Gamma dt}\\
&\sim \tilde c(t) - 2 \Gamma \tilde c(t) dt ,
\end{split}
\end{equation}
where $n_{eq}(t)=n_{eq}(\omega_k(t))$ is the mean occupation number when the system is in equilibrium, typically the thermal distribution of Eq.~\eqref{eq:thermal}.
When coupled to such incoherent states, the equilibrium value of the coherence parameter $c_{eq}$ vanishes, as is assumed in Eqs.~\eqref{eq:infdiss}.

To first order in $dt$, Eqs.~\eqref{eq:infbog}-\eqref{eq:infdiss} combine to give
\begin{equation}
\label{eq:eomncdissip}
\begin{split}
(\partial_t+ 2 \Gamma)n &= 2 \Gamma n_{eq} + 2 \Re \left [ ( \alpha^* \partial_t \beta -\beta \partial_t \alpha^* ) c \right ] , \\
(\partial_t+ 2 \Gamma)c &= ( \alpha \partial_t \beta^* -\beta^* \partial_t \alpha ) (1+2n) \\
&+ 2 (\alpha^* \partial_t \alpha - \beta \partial_t \beta^*) c .
\end{split}
\end{equation}
These are equivalent to Eqs.~\eqref{eq:eomdissipncexact} when the non-dissipative equations for $\alpha(t)$ and $\beta(t)$, Eqs.~\eqref{eq:eomalphabeta}, are substituted in Eqs.~\eqref{eq:eomncdissip}, and the unitarity condition $\abs{\alpha}^2 -\abs{\beta}^2 =1$ is used. One can check that in the limit $\Gamma \to 0$, Eqs.~\eqref{eq:outnc} satisfy the above equations. One can also check that, when there is no modulation or after the modulation has ended -- i.e., when $\alpha$, $\beta$ and $n_{eq}$ are constant in time -- Eqs.~\eqref{eq:eomncdissip} imply that $n$ and $c$ decay exponentially toward their equilibrium values $n_{eq}$ and $0$. 

Furthermore, Eqs.~\eqref{eq:eomncdissip} yield the following simple equation for the evolution of the effective number $\bar n = n(n+1) - \abs{c}^2$ that fixes the value of the entropy~\cite{Campo:2005sy}:
\begin{equation}
\begin{split}
\partial_t \bar n = 2 \Gamma \left [ 2 \abs{c}^2 + (1+2n)(n_{eq} - n) \right ] .
\end{split}
\end{equation}
The evolution of $\bar n(t)$ determined by this equation governs the entropy exchanges between the system and its environment.

\bibliographystyle{../../biblio/nbrevtitle}
\bibliography{../../biblio/bibliopubli}

\end{document}